\documentclass{article}
\usepackage[applemac]{inputenc}
\usepackage{cite}
\usepackage{amssymb}
\usepackage{amsmath}
\usepackage{graphicx}
\usepackage{relate}
\usepackage{a4}
\usepackage{fullpage}

\newcommand{\ceil}[1]{\left\lceil{#1}\right\rceil}

\newtheorem{lemma}{Lemma}
\newtheorem{theorem}{Theorem}

\newcommand{\qed}{\hfill\ensuremath{\Box}\medskip\\\noindent}
\newenvironment{proof}{\noindent\emph{Proof. }}

\newcommand{\prcfont}[1]{{\ensuremath{\mathsf{#1}}}}

\newcommand{\Move}{\prcfont{Move}}
\newcommand{\Close}{\prcfont{Close}}

\newcommand{\Insert}{\prcfont{Insert}}

\newcommand{\Member}{\prcfont{Member}}

\title{New Algorithms for Regular Expression Matching}

\author{Philip Bille\thanks{The IT University of Copenhagen, Rued Langgaards Vej 7, 2300
Copenhagen S, Denmark. Email: {\tt beetle@itu.dk}. An extended abstract of this paper appeared in Proceedings of the 33rd International Colloquium on Automata, Languages and Programming, 2006.}}
\date{\today}

\begin{document}
\maketitle

\begin{abstract}
In this paper we revisit the classical regular expression matching problem, namely, given a regular expression $R$  and a string $Q$, decide if $Q$ matches one of the strings specified by $R$. Let $m$ and $n$ be the length of $R$ and $Q$, respectively. On a standard unit-cost RAM with word length $w \geq \log n$, we show that the problem can be solved in $O(m)$ space with the following running times:
\begin{equation*}
\begin{cases}
      O(n\frac{m \log w}{w} + m \log w) & \text{ if $m > w$} \\
      O(n\log m + m\log m) & \text{ if $\sqrt{w} < m \leq  w$} \\
      O(\min(n+ m^2, n\log m + m\log m)) & \text{ if $m \leq \sqrt{w}$.}
\end{cases}
\end{equation*} 
This improves the best known time bound among algorithms using $O(m)$ space. Whenever $w \geq \log^2 n$ it improves all known time bounds regardless of how much space is used. 
\end{abstract}

\section{Introduction} 
Regular expressions are a powerful and simple way to describe a set of strings. For this reason, they are often chosen as the input language for text processing applications. For instance, in the lexical analysis phase of compilers, regular expressions are often used to specify and distinguish tokens to be passed to the syntax analysis phase. Utilities such as Grep, the programming language Perl, and most modern text editors provide mechanisms for handling regular expressions. These applications all need to solve the classical \textsc{Regular Expression Matching} problem, namely, given a regular expression $R$ and a string $Q$, decide if $Q$ matches one of the strings specified by $R$. 

The standard textbook solution, proposed by Thompson~\cite{Thomp1968} in 1968, constructs a \emph{non-deterministic finite automaton} (NFA) accepting all strings matching $R$. Subsequently, a state-set simulation checks if the NFA accepts $Q$. This leads to a simple $O(nm)$ time and $O(m)$ space algorithm, where $m$ and $n$ are the number of symbols in $R$ and $Q$, respectively. The full details are reviewed later in Sec.~\ref{sec:regex} and can found in most textbooks on compilers (e.g. Aho et. al. \cite{ASU1986}). Despite the importance of the problem, it took 24 years before the $O(nm)$ time bound was improved by Myers~\cite{Myers1992} in 1992, who achieved $O(\frac{nm}{\log n} + (n+m)\log n)$ time and $O(\frac{nm}{\log n})$ space. For most values of $m$ and $n$ this improves the $O(nm)$ algorithm by a $O(\log n)$ factor. Currently, this is the fastest known algorithm. Recently, Bille and Farach-Colton~\cite{BFC2005} showed how to reduce the space of Myers' solution to $O(n)$. Alternatively, they showed how to achieve a speedup of $O(\log m)$ over Thompson's algorithm while using $O(m)$ space. These results are all valid on a unit-cost RAM with $w$-bit words and a standard instruction set including addition, bitwise boolean operations, shifts, and multiplication. Each word is capable of holding a character of $Q$ and hence $w \geq \log n$. The space complexities refer to the number of words used by the algorithm, not counting the input which is assumed to be read-only. All results presented here assume the same model. In this paper we present new algorithms achieving the following complexities:
\begin{theorem}\label{thm:main}
Given a regular expression $R$ and a string $Q$ of lengths $m$ and $n$, respectively, \textsc{Regular Expression Matching} can be solved using $O(m)$ space with the following running times:
\begin{equation*}
\begin{cases}
      O(n\frac{m \log w}{w} + m \log w) & \text{ if $m > w$} \\
      O(n\log m + m\log m) & \text{ if $\sqrt{w} < m \leq  w$} \\
      O(\min(n+ m^2, n\log m + m\log m)) & \text{ if $m \leq \sqrt{w}$.}
\end{cases}
\end{equation*} 
\end{theorem}
This represents the best known time bound among algorithms using $O(m)$ space. To compare these with previous results, consider a conservative word length of $w = \log n$. When the regular expression is "large", e.g.,  $m > \log n$, we achieve an $O(\frac{\log n}{\log \log n})$ factor speedup over Thompson's algorithm using $O(m)$ space. Hence, we simultaneously match the best known time and space bounds for the problem, with the exception of an $O(\log \log n)$ factor in time. More interestingly, consider the case when the regular expression is "small", e.g., $m = O(\log n)$. This is usually the case in most applications. To beat the $O(n\log n)$ time of Thompson's algorithm, the fast algorithms~\cite{Myers1992, BFC2005} essentially convert the NFA mentioned above into a \emph{deterministic finite automaton} (DFA) and then simulate this instead. Constructing and storing the DFA incurs an additional exponential time and space cost in $m$, i.e., $O(2^m) = O(n)$. However, the DFA can now be simulated in $O(n)$ time, leading to an $O(n)$ time and space algorithm. Surprisingly, our result shows that this exponential blow-up in $m$ can be avoided with very little loss of efficiency. More precisely, we get an algorithm using $O(n\log \log n)$ time and $O(\log n)$ space. Hence, the space is improved exponentially at the cost of an $O(\log \log n)$ factor in time. In the case of an even smaller regular expression, e.g., $m = O(\sqrt{\log n})$, the slowdown can be eliminated and we achieve optimal $O(n)$ time. For larger word lengths our time bounds improve. In particular, when $w > \log n \log \log n$ the bound is better in all cases, except for $\sqrt{w} \leq m \leq w$, and when $w > \log^2n$ it improves all known time bounds regardless of how much space is used.

The key to obtain our results is to avoid explicitly converting small NFAs into DFAs. Instead we show how to effectively simulate them directly using the parallelism available at the word-level of the machine model. The kind of idea is not new and has been applied to many other string matching problems, most famously, the Shift-Or algorithm~\cite{BYG1992}, and the approximate string matching algorithm by Myers \cite{Myers1999}. However, none of these algorithms can be easily extended to \textsc{Regular Expression Matching}. The main problem is the complicated dependencies between states in an NFA. Intuitively, a state may have long paths of $\epsilon$-transitions to a large number of other states, all of which have to be traversed in parallel in the state-set simulation. To overcome this problem we develop several new techniques ultimately leading to Theorem~\ref{thm:main}. For instance, we introduce a new hierarchical decomposition of NFAs suitable for a parallel state-set simulation. We also show how state-set simulations of large NFAs efficiently reduces to simulating small NFAs.

The results presented in this paper are primarily of theoretical interest. However, we believe that most of the ideas are useful in practice. The previous algorithms require large tables for storing DFAs, and perform a long series of lookups in these tables. As the tables become large we can expect a high number of cache-misses during the lookups, thus limiting the speedup in practice. Since we avoid these tables, our algorithms do not suffer from this defect. 

The paper is organized as follows. In Sec.~\ref{sec:regex} we review Thompson's NFA construction, and in Sec.~\ref{sec:simul} we present the above mentioned reduction. In Sec.~\ref{sec:simple} we present our first simple algorithm for the problem which is then improved in Sec.~\ref{sec:notsimple}. Combining these algorithms with our reduction leads to Theorem~\ref{thm:main}. We conclude with a couple of remarks and open problems in Sec.~\ref{sec:remarks}.

\section{Regular Expressions and Finite Automata}\label{sec:regex}
In this section we briefly review Thompson's construction and the standard
state-set simulation.  The set of \emph{regular expressions} over an alphabet 
$\Sigma$ are defined recursively as follows: 
\begin{itemize}
\item A character $\alpha \in \Sigma$ is a regular expression.
\item If $S$ and $T$ are regular expressions then so is the
  \emph{concatenation}, $(S)\cdot(T)$, the \emph{union}, $(S)|(T)$, and
  the \emph{star}, $(S)^*$.
\end{itemize}
Unnecessary parentheses can be removed by observing that $\cdot$ and
$|$ is associative and by using the standard precedence of the
operators, that is $*$ precedes $\cdot$, which in turn precedes $|$. We often remove the $\cdot$ when writing regular expressions. 

\begin{figure}[t] 
  \centering \includegraphics[scale=.5]{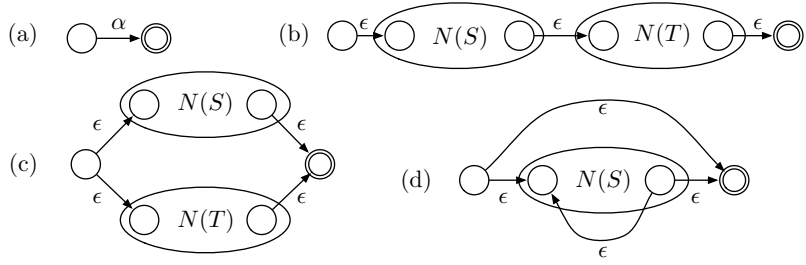}
  \caption{Thompson's NFA construction. The regular expression for a character $\alpha \in \Sigma$ corresponds to NFA $(a)$. If $S$ and $T$ are regular expressions then $N(ST)$, $N(S|T)$, and
    $N(S^*)$ correspond to NFAs $(b)$, $(c)$, and $(d)$, respectively.
    Accepting nodes are marked with a double circle.}
  \label{fig:thompson}
\end{figure}

The \emph{language} $L(R)$ generated by $R$ is the set of all strings matching $R$. The \emph{parse tree} $T(R)$ of $R$ is the binary rooted tree representing the hiearchical structure of $R$. Each leaf is labeled by a character in $\Sigma$ and each internal node is labeled either $\cdot$, $|$, or $*$. A \emph{finite automaton} is a tuple $A = (V, E, \delta, \theta, \phi)$, where 
\begin{itemize}
  \item $V$ is a set of nodes called \emph{states},
  \item $E$ is set of directed edges between states called \emph{transitions}, 
  \item $\delta : E \rightarrow \Sigma \cup \{\epsilon\}$ is a function assigning labels to transitions, and
  \item $\theta, \phi \in V$ are distinguished states called the \emph{start state} and \emph{accepting state}, respectively\footnote{Sometimes NFAs are allowed a \emph{set} of accepting states, but this is not necessary for our purposes.}.
\end{itemize} 
Intuitively, $A$ is an edge-labeled directed graph with special start and accepting nodes. $A$ is a \emph{deterministic finite automaton} (DFA) if $A$ does not contain any $\epsilon$-transitions, and all outgoing transitions of any state have different labels. Otherwise, $A$ is a \emph{non-deterministic automaton} (NFA). We say that $A$ \emph{accepts} a string $Q$ if there is a path from $\theta$ to $\phi$ such that the concatenation of labels on the path spells out $Q$. 
Thompson~\cite{Thomp1968} showed how to recursively construct a NFA $N(R)$ accepting all strings in $L(R)$. 
The rules are presented below and illustrated in Fig.~\ref{fig:thompson}.

\begin{itemize}
\item $N(\alpha)$ is the automaton consisting of states $\theta_\alpha$, $\phi_\alpha$, and an $\alpha$-transition
  from $\theta_\alpha$ to $\phi_\alpha$.
\item Let $N(S)$ and $N(T)$ be automata for regular expressions $S$ and
  $T$ with start and accepting states $\theta_S$, $\theta_T$, $\phi_S$,
  and $\phi_T$, respectively. Then, NFAs $N(S\cdot T)$, $N(S|T)$,
  and $N(S^*)$ are constructed as follows:
  \begin{relate}
  \item[$N(ST)$:] Add start state $\theta_{ST}$ and accepting state $\phi_{ST}$, and $\epsilon$-transitions $(\theta_{ST}, \theta_{S})$, $(\phi_{S}, \theta_T)$, and $(\phi_T, \phi_{ST})$.
  \item[$N(S|T)$:] Add start state $\theta_{S|T}$ and accepting state $\phi_{S|T}$, and add $\epsilon$-transitions $(\theta_{S|T}, \theta_S)$, $(\theta_{S|T}, \theta_T)$, $(\phi_S, \phi_{S|T})$, and $(\phi_T, \phi_{S|T})$.
  \item [$N(S^*)$:] Add a new start state $\theta_{S^*}$ and
    accepting state $\phi_{S^*}$, and $\epsilon$-transitions $(\theta_{S^*}, \theta_S)$, $(\theta_{S^*}, \phi_{S^*})$, $(\phi_S, \phi_{S^*})$, and $(\phi_S, \theta_S)$.
\end{relate}
\end{itemize}

Readers familiar with Thompson's construction will notice that $N(ST)$ is slightly different from the usual construction. 
This is done to simplify our later presentation and does not affect the worst case complexity of the problem. Any automaton produced by these rules we call a \emph{Thompson-NFA} (TNFA). By construction, $N(R)$ has a single start and accepting state, denoted $\theta$ and $\phi$, respectively. $\theta$ has no incoming transitions and $\phi$ has no outgoing transitions. The total number of states is $2m$ and since each state has at most $2$ outgoing transitions that the total number of transitions is at most $4m$.  Furthermore, all incoming
transitions have the same label, and we denote a state with incoming $\alpha$-transitions an \emph{$\alpha$-state}. Note that the star construction in Fig. \ref{fig:thompson}(d) introduces a transition from the accepting state of $N(S)$ to the start state of $N(S)$. All such transitions are called \emph{back transitions} and all other transitions are \emph{forward transitions}. We need the following property. 
\begin{lemma}[Myers \cite{Myers1992}]\label{lem:cyclefree} Any cycle-free path in a TNFA contains at most one back transition.
\end{lemma}
For a string $Q$ of length $n$ the standard state-set simulation of
$N(R)$ on $Q$ produces a sequence of state-sets $S_0, \ldots, S_n$. The
$i$th set $S_i$, $0\leq i \leq n$, consists of all states in $N(R)$ for which there is a path from $\theta$ that spells out the $i$th prefix of $Q$. The simulation can be implemented with the following simple
operations. For a state-set $S$ and a character $\alpha \in \Sigma$, define
\begin{relate}
\item[$\Move(S,\alpha)$:] Return the set of states reachable from $S$ via a single $\alpha$-transition.
\item[$\Close(S)$:] Return the set of states reachable from $S$ via $0$ or more $\epsilon$-transitions.
\end{relate}
Since the number of states and transitions in $N(R)$ is $O(m)$, both operations can be easily implemented in $O(m)$ time. The $\Close$ operation is often called an \emph{$\epsilon$-closure}. The simulation proceeds as follows: Initially, $S_0 := \Close(\{\theta\})$. If $Q[j]=\alpha$, $1\leq j \leq
n$, then $S_j := \Close(\Move(S_{j-1}, \alpha))$. Finally, $Q \in L(R)$ iff $\phi \in S_n$. Since each state-set $S_j$ only depends on $S_{j-1}$ this algorithm uses $O(mn)$ time and $O(m)$ space. 

\section{From Large to Small TNFAs}\label{sec:simul}
In this section we show how to simulate $N(R)$ by simulating a number of smaller TNFAs. We will use this to achieve our bounds when $R$ is large. 

\subsection{Clustering Parse Trees and Decomposing TNFAs}\label{sec:clustering}
\begin{figure}[t]
  \centering \includegraphics[scale=.5]{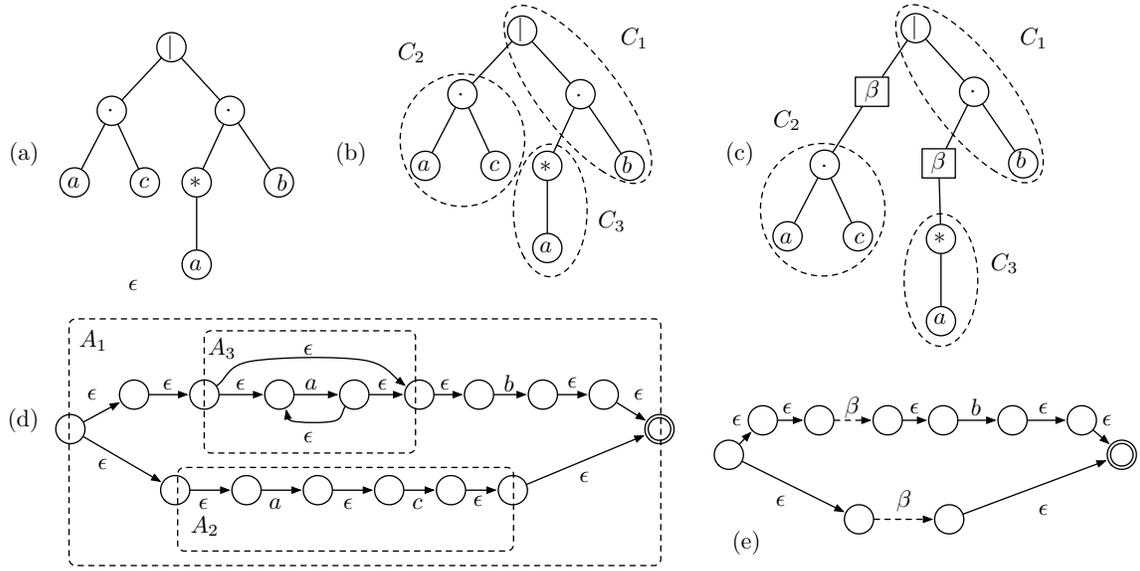}
   \caption{(a) The parse tree for the regular expression
     $ac|a^*b$. (b) A clustering of $(a)$ into
     node-disjoint connected subtrees $C_1$, $C_2$, and $C_3$, each with at most $3$ nodes. (c) The clustering from (b) extended with pseudo-nodes. (d) The nested decomposition of $N(ac|a^*b)$. (e) The TNFA corresponding to $C_1$.}
   \label{fig:clustering}
\end{figure}

Let $R$ be a regular expression of length $m$. We first show how to decompose $N(R)$ into smaller TNFAs. This decomposition is based on a simple clustering of the parse tree $T(R)$. A \emph{cluster} $C$ is a connected subgraph of $T(R)$ and a \emph{cluster partition} $CS$ is a partition of the nodes of $T(R)$ into node-disjoint clusters. Since $T(R)$ is a binary tree with $O(m)$ nodes, a simple top-down procedure provides the following result (see e.g. \cite{Myers1992}):
\begin{lemma}\label{lem:clustering}
Given a regular expression $R$ of length $m$ and a parameter $x$, a cluster partition $CS$ of $T(R)$ can be constructed in $O(m)$ time such that $|CS| = O(\ceil{m/x})$, and for any $C\in CS$, the number of nodes in $C$ is
  at most $x$.
\end{lemma}
For a cluster partition $CS$, edges adjacent to two clusters are \emph{external edges} and all other edges are \emph{internal edges}. Contracting all internal edges in $CS$ induces a \emph{macro tree}, where each cluster is represented by a single \emph{macro node}. Let $C_v$ and $C_w$ be two clusters with corresponding macro nodes $v$ and $w$. We say that $C_v$ is the \emph{parent cluster} (resp. \emph{child cluster}) of $C_w$ if $v$ is the parent (resp. child) of $w$ in the macro tree. The \emph{root cluster and leaf clusters} are the clusters corresponding to the root and the leaves of the macro tree. An example clustering of a parse tree is shown in Fig.~\ref{fig:clustering}(b).
Given a cluster partition $CS$ of $T(R)$ we show how to divide $N(R)$ into a set of small nested TNFAs. Each cluster $C \in CS$ will correspond to a TNFA $A$, and we use the terms child, parent, root, and leaf for the TNFAs in the same way we do with clusters. For a cluster $C \in CS$ with children $C_1, \ldots, C_l$, insert  a special \emph{pseudo-node} $p_i$, $1 \leq i \leq l$, in the middle of the external edge connecting $C$ with $C_i$. We label each pseudo-node by a special character $\beta \not \in \Sigma$. Let $T_C$ be the tree induced by the set of nodes in $C$ and $\{p_1, \ldots, p_l\}$. Each leaf in $T_C$ is labeled with a character from $\Sigma \cup \{\beta\}$, and hence $T_C$ is a well-formed parse tree for some regular expression $R_C$ over $\Sigma \cup \{\beta\}$. Now, the TNFA $A$ corresponding to $C$ is $N(R_C)$. In $A$, child TNFA $A_i$ is represented by its start and accepting state $\theta_{A_i}$ and $\phi_{A_i}$ and a \emph{pseudo-transition} labeled $\beta$ connecting them. An example of these definitions is given in Fig.~\ref{fig:clustering}. We call any set of TNFAs obtained from a cluster partition as above a \emph{nested decomposition} $AS$ of $N(R)$. 
\begin{lemma}\label{lem:decomp}
  Given a regular expression $R$ of length $m$ and a parameter $x$, a nested decomposition $AS$ of $N(R)$ can be constructed in $O(m)$ time such that $|AS| = O(\ceil{m/x})$, and for any $A\in AS$, the number of states in $A$ is at most $x$.
\end{lemma}
\begin{proof}
Construct the parse tree $T(R)$ for $R$ and build a cluster partition $CS$ according to Lemma~\ref{lem:clustering} with parameter $y = \frac{x}{4} - \frac{1}{2}$. From $CS$ build a nested decomposition $AS$ as described above. Each $C \in CS$ corresponds to a TNFA $A \in AS$ and hence $|AS| = O(\ceil{m/y}) = O(\ceil{m/x})$. Furthermore, if $|V(C)| \leq y$ we have $|V(T_C)| \leq 2y + 1$. Each node in $T_C$ contributes two states to the corresponding TNFA $A$, and hence the total number of states in $A$ is at most $4y + 2 = x$. Since the parse tree, the cluster partition, and the nested decomposition can be constructed in $O(m)$ time the result follows.\qed
\end{proof}

\subsection{Simulating Large Automata}
We now show how $N(R)$ can be simulated using the TNFAs in a nested decomposition. For this purpose we define a simple data structure to dynamically maintain the TNFAs. Let $AS$ be a nested decomposition of $N(R)$ according to Lemma~\ref{lem:decomp}, for some parameter $x$. Let $A \in AS$ be a TNFA, let $S_A$ be a state-set of $A$, let $s$ be a state in $A$, and let $\alpha \in \Sigma$.  A \emph{simulation data structure} supports the $4$ operations: $\Move_A(S_A, \alpha)$, $\Close_A(S_A)$, $\Member_A(S_A, s)$, and $\Insert_A(S_A, s)$. Here, the operations $\Move_A$ and $\Close_A$ are defined exactly as in Sec.~\ref{sec:regex}, with the modification that they only work on $A$ and not $N(R)$. The operation $\Member_A(S_A, s)$ returns yes if $s \in S_A$ and no otherwise and $\Insert_A(S_A, s)$ returns the set $S_A \cup \{s\}$.

In the following sections we consider various efficient implementations of simulation data structures. For now assume that we have a black-box data structure for each $A \in AS$. To simulate $N(R)$ we proceed as follows. First, fix an ordering of  the TNFAs in the nested decomposition $AS$, e.g., by a preorder traversal of the tree represented given by the parent/child relationship of the TNFAs. The collection of state-sets for each TNFA in $AS$ are represented in a \emph{state-set array} $X$ of length $|AS|$. The state-set array is indexed by the above numbering, that is, $X[i]$ is the state-set of the $i$th TNFA in $AS$. For notational convenience we write $X[A]$ to denote the entry in $X$ corresponding to $A$. Note that a parent TNFA share two states with each child, and therefore a state may be represented more than once in $X$. To avoid complications we will always assure that $X$ is \emph{consistent}, meaning that if a state $s$ is included in the state-set of some TNFA, then it is also included in the state-sets of all other TNFAs that share $s$. If $S =  \bigcup_{A\in AS} X[A]$ we say that $X$ \emph{models} the state-set $S$ and write $S \equiv X$. 

Next we show how to do a state-set simulation of $N(R)$ using the operations $\Move_{AS}$ and $\Close_{AS}$, which we define below. These operations recursively update a state-set array using the simulation data structures. For any $A\in AS$, state-set array $X$, and $\alpha \in \Sigma$ define
\begin{relate}
\item[$\Move_{AS}(A, X, \alpha)$:] 
\begin{enumerate}
\item $X[A] :=  \Move_{A}(X[A], \alpha)$
\item For each child $A_i$ of $A$ in topological order do
\begin{enumerate}
\item $X := \Move_{AS}(A_i, X, \alpha)$
\item If $\phi_{A_i} \in X[A_i]$ then $X[A] := \Insert_A(X[A],\phi_{A_i})$
\end{enumerate}
\item Return $X$
\end{enumerate}
\item[$\Close_{AS}(A, X)$:] 
\begin{enumerate}
\item $X[A] := \Close_A(X[A])$
\item For each child $A_i$ of $A$ in topological order do
\begin{enumerate}
\item If $\theta_{A_i} \in X[A]$ then $X[A_i] := \Insert_{A_i}(X[A_i], \theta_{A_i})$
\item X := $\Close_{AS}(A_i, X)$
\item If $\phi_{A_i} \in X[A_i]$ then $X[A] := \Insert_A(X[A],\phi_{A_i})$
\item $X[A] := \Close_A(X[A])$
\end{enumerate}
\item Return $X$
\end{enumerate}
\end{relate}
The $\Move_{AS}$ and $\Close_{AS}$ operations recursively traverses the nested decomposition top-down processing the children in topological order. At each child the shared start and accepting states are propagated in the state-set array. For simplicity, we have written $\Member_A$ using the symbol $\in$.

The state-set simulation of $N(R)$ on a string $Q$ of length $n$ produces the sequence of state-set arrays $X_0, \ldots, X_n$ as follows: Let $A_r$ be the root automaton and let $X$ be an empty state-set array (all entries in $X$ are $\emptyset$). Initially, set $X[A_r] := \Insert_{A_r}(X[A_r], \theta_{A_r})$ and compute $X_0 := \Close_{AS}(A_r, \Close_{AS}(A_r, X))$. For $i>0$ we compute $X_{i}$ from $X_{i-1}$ as follows: 
\begin{equation*}
X_i := \Close_{AS}(A_r, \Close_{AS}(A_r, \Move_{AS}(A_r, X_{i-1}, Q[i])))
\end{equation*}
Finally, we output $Q \in L(R)$ iff $\phi_{A_r} \in X_n[A_r]$. To see that this algorithm correctly solves \textsc{Regular Expression Matching} it suffices to show that for any $i$, $0\leq i \leq n$, $X_i$ correctly models the $i$th state-set $S_i$ in the standard state-set simulation.  We need the following lemma. 
\begin{lemma}\label{lem:statearray}
Let $X$ be a state-set array and let $A_r$ be the root TNFA in a nested decomposition $AS$. If $S$ is the state-set modeled by $X$, then  
\begin{itemize}
  \item $\Move(S,\alpha) \equiv \Move_{AS}(A_r, X, \alpha)$ and 
  \item $\Close(S) \equiv \Close_{AS}(A_r, \Close_{AS}(A_r, X))$.
\end{itemize}
\end{lemma}
\begin{proof}
First consider the $\Move_{AS}$ operation. Let $\overline{A}$ be the TNFA induced by all states in $A$ and descendants of $A$ in the nested decomposition,  i.e., $\overline{A}$ is obtained by recursively "unfolding" the pseudo-states and pseudo-transitions in $A$, replacing them by the TNFAs they represent.  We show by induction that the state-array $X_A' := \Move_{AS}(A, X, \alpha)$ models $\Move(S,\alpha)$ on $\overline{A}$. In particular, plugging in $A = A_r$, we have that $\Move_{AS}(A_r, X, \alpha)$ models $\Move(S,\alpha)$ as required. 

Initially, line $1$ updates $X[A]$ to be the set of states reachable from a single $\alpha$-transition in $A$.
If $A$ is a leaf, line $2$ is completely bypassed and the result follows immediately. Otherwise, let $A_1, \ldots, A_l$ be the children of $A$ in topological order. Any incoming transition to a state $\theta_{A_i}$ or outgoing transition from a state $\phi_{A_i}$ is an $\epsilon$-transition by Thompson's construction. Hence, no endpoint of an $\alpha$-transition in $A$ can be shared with any of the children $A_1, \ldots, A_l$. It follows that after line $1$ the updated $X[A]$ is the desired state-set, except for the shared states, which have not been handled yet. By induction, the recursive calls in line $2$(a) handle the children. Among the shared states only the accepting ones, $\phi_{A_1}, \ldots, \phi_{A_l}$, may be the endpoint of an $\alpha$-transition and therefore line $2$(b) computes the correct state-set.

The $\Close_{AS}$ operation proceeds in a similar, though slightly more complicated fashion. Let $\widetilde{X}_A$ be the state-array modeling the set of states reachable via a path of \emph{forward} $\epsilon$-transitions in $\overline{A}$, and let $\widehat{X}_A$ be the state array modelling $\Close(S)$ in $\overline{A}$. We show by induction that if $X_A'' := \Close_{AS}(A, X)$ then
\begin{equation*}
\widetilde{X}_A \subseteq X_A'' \subseteq \widehat{X}_A,
\end{equation*}
where the inclusion refers to the underlying state-sets modeled by the state-set arrays. Initially, line $1$ updates $X[A] := \Close_{A}(X[A])$. If $A$ is a leaf then clearly $X_A'' = \widehat{X}_A$. Otherwise, let $A_1, \ldots, A_l$ be the children of $A$ in topological order. Line $2$ recursively update the children and propagate the start and accepting states in (a) and (c). Following each recursive call we again update $X[A] := \Close_{A}(X[A])$ in (d). No state is included in $X_A''$ if there is no $\epsilon$-path in $A$ or through any child of $A$. Furthermore, since the children are processed in topological order it is straightforward to verify that the sequence of updates in line $2$ ensure that $X_A''$ contain all states reachable via a path of forward $\epsilon$-transitions in $A$ or through a child of $A$. Hence, by induction we have $\widetilde{X}_A \subseteq X_A'' \subseteq \widehat{X}_A$ as desired. 

A similar induction shows that the state-set array $\Close_{AS}(A_r, X'')$ models the set of states reachable from $X''$ using a path consisting of forward $\epsilon$-transitions and at most $1$ back transition. However, by Lemma~\ref{lem:cyclefree} this is exactly the set of states reachable by a path of $\epsilon$-transitions. Hence, $\Close_{AS}(A_r, X'')$ models $\Close(S)$ and the result follows.\qed
\end{proof}

By Lemma~\ref{lem:statearray} the state-set simulation can be done using the $\Close_{AS}$ and $\Move_{AS}$ operations and the complexity now directly depends on the complexities of the simulation data structure. Putting it all together the following reduction easily follows:
\begin{lemma}\label{lem:simulation}
Let $R$ be a regular expression  of length $m$ over alphabet $\Sigma$ and let $Q$ a string of length $n$. Given a simulation data structure for TNFAs with $x < m$ states over alphabet $\Sigma \cup \{\beta\}$, where $\beta \not\in \Sigma$, that supports all operations in $O(t(x))$ time, using $O(s(x))$ space, and $O(p(x))$ preprocessing time, \textsc{Regular Expression Matching} for $R$ and $Q$ can be solved in $O(\frac{nm \cdot t(x)}{x} + \frac{m\cdot p(x)}{x})$ time using $O(\frac{m \cdot s(x)}{x})$ space. 
\end{lemma}
\begin{proof}
Given $R$ first compute a nested decomposition $AS$ of $N(R)$ using Lemma~\ref{lem:decomp} for parameter $x$. For each TNFA $A \in AS$ sort $A$'s children to topologically and keep pointers to start and accepting states. By Lemma~\ref{lem:decomp} and since topological sort can be done in $O(m)$ time this step uses $O(m)$ time. The total space to represent the decomposition is $O(m)$. Each $A \in AS$ is a TNFA over the alphabet $\Sigma \cup \{\beta\}$ with at most $x$ states and $|AS| = O(\frac{m}{x})$. Hence, constructing simulation data structures for all $A \in AS$ uses $O(\frac{m p(x)}{x})$ time and $O(\frac{m s(x)}{x})$ space. With the above algorithm the state-set simulation of $N(R)$ can now be done in $O(\frac{nm \cdot t(x)}{x})$ time, yielding the desired complexity.\qed
\end{proof}

The idea of decomposing TNFAs is also present in Myers' paper~\cite{Myers1992}, though he does not give a "black-box" reduction as in Lemma~\ref{lem:simulation}. We believe that the framework provided by Lemma~\ref{lem:simulation} helps to simplify the presentation of the algorithms significantly. We can restate Myers' result in our setting as the existence of a simulation data structure with $O(1)$ query time that uses $O(x\cdot 2^x)$ space and preprocessing time. For $x \leq \log (n/\log n)$ this achieves the result mentioned in the introduction. The key idea is to encode and tabulate the results of all queries (such an approach is frequently referred to as the "Four Russian Technique"~\cite{ADKF1970}). Bille and Farach~\cite{BFC2005} give a more space-efficient encoding that does not use Lemma~\ref{lem:simulation} as above. Instead they show how to encode \emph{all possible} simulation data structures in total $O(2^x + m)$ time and space while maintaining $O(1)$ query time.  

In the following sections we show how to efficiently avoid the large tables needed in the previous approaches. Instead we implement the operations of simulation data structures using the word-level parallelism of the machine model. 

\section{A Simple Algorithm}\label{sec:simple}
In this section we present a simple simulation data structure for TNFAs, and develop some of the ideas for the improved result of the next section. Let $A$ be a TNFA with $m = O(\sqrt{w})$ states. We will show how to support all operations in $O(1)$ time using $O(m)$ space and $O(m^2)$ preprocessing time.

To build our simulation data structure for $A$, first sort all states in $A$ in topological order ignoring the back transitions.  We require that the endpoints of an $\alpha$-transition are consecutive in this order. This is automatically guaranteed using a standard $O(m)$ time algorithm for topological sorting (see e.g. \cite{CLRS2001}). We will refer to states in $A$ by their rank in this order. A state-set of $A$ is represented using a bitstring $S = s_1s_2\ldots s_m$ defined such that $s_i = 1$ iff node $i$ is in the state-set. The simulation data structure consists of the following bitstrings:
\begin{itemize}
\item For each $\alpha \in \Sigma$,  a string $D_\alpha = d_1 \ldots d_m$ such that $d_i = 1$ iff $i$ is an $\alpha$-state. 
\item A string $E = 0e_{1,1}e_{1,2}\ldots e_{1,m}0e_{2,1}e_{2,2}\ldots e_{2,m}0 \ldots 0 e_{m,1}e_{m,2}\ldots e_{m,m}$, where $e_{i,j} = 1$ iff $i$ is $\epsilon$-reachable from $j$. The zeros are \emph{test bits} needed for the algorithm.
\item Three constants $I = (10^m)^m$, $X = 1(0^m1)^{m-1}$, and $C = 1(0^{m-1}1)^{m-1}$. Note that $I$ has a $1$ in each test bit position\footnote{We use exponentiation to denote repetition, i.e., $1^30 = 1110$.}.
\end{itemize}
The strings $E$, $I$, $X$, and $C$ are easily computed in $O(m^2)$ time and use $O(m^2)$ bits. Since $m = O(\sqrt{w})$ only $O(1)$ space is needed to store these strings. We store $D_\alpha$ in a hashtable indexed by $\alpha$. Since the total number of different characters in $A$ can be at most $m$, the hashtable contains at most $m$ entries. Using perfect hashing $D_\alpha$ can be represented in $O(m)$ space with $O(1)$ worst-case lookup time. The preprocessing time is expected $O(m)$ w.h.p.. To get a worst-case bound we use the deterministic dictionary of Hagerup et. al. \cite{HMP2001} with $O(m\log m)$ worst-case preprocessing time. In total the data structure requires $O(m)$ space and $O(m^2)$ preprocessing time. 

Next we show how to support each of the operations on $A$. Suppose $S = s_1 \ldots s_m$ is a bitstring representing a state-set of $A$ and $\alpha \in \Sigma$. The result of $\Move_A(S,\alpha)$ is given by
\begin{equation*}
S' := (S >> 1) \: \& \: D_\alpha.
\end{equation*}
This should be understood as C notation, where the right-shift is unsigned. Readers familiar with the Shift-Or algorithm~\cite{BYG1992} will notice the similarity. To see the correctness, observe that state $i$ is put in $S'$ iff state $(i-1)$ is in $S$ and the $i$th state is an $\alpha$-state. Since the endpoints of $\alpha$-transitions  are consecutive in the topological order it follows that $S'$ is correct. Here, state $(i-1)$ can only influence state $i$, and this makes the operation easy to implement in parallel. However, this is not the case for $\Close_A$. Here, any state can potentially affect a large number of states reachable through long $\epsilon$-paths. To deal with this we use the following steps.
\begin{align*}
Y &:= (S \times X) \:\&\: E \\
Z &:= ((Y \: |\: I) - (I >> m)) \:\&\: I \\
S' &:= ((Z \times C) << w-m(m+1)) >> w - m
\end{align*}
We describe in detail why this, at first glance somewhat cryptic sequence, correctly computes $S'$ as the result of $\Close_A(S)$. The variables $Y$ and $Z$ are simply temporary variables inserted to increase the readability of the computation. Let $S = s_1 \ldots s_m$. Initially, $S \times X$ concatenates $m$ copies of $S$ with a zero bit between each copy, that is, 
\begin{equation*}
S \times X = s_1\ldots s_m \times 1(0^m1)^{m-1} = (0s_1\ldots s_m)^m.
\end{equation*}
The bitwise $\&$ with $E$ gives 
\begin{equation*}
Y = 0y_{1,1}y_{1,2}\ldots y_{1,m}0y_{2,1}y_{2,2}\ldots y_{2,m}0 \ldots 0 y_{m,1}y_{m,2}\ldots y_{m,m},
\end{equation*} 
where $y_{i,j} = 1$ iff state $j$ is in $S$ and state $i$ is $\epsilon$-reachable from $j$. In other words, the substring $Y_i = y_{i,1} \ldots y_{i,m}$ indicates the set of states in $S$ that have a path of $\epsilon$-transitions to $i$. Hence, state $i$ should be included in $\Close_A(S)$ precisely if at least one of the bits in $Y_i$ is $1$. This is determined next. First $(Y \: | \: I) - (I >> m)$ sets all test bits to $1$ and subtracts the test bits shifted right by $m$ positions. This ensures that if all positions in $Y_i$ are $0$, the $i$th test bit in the result is $0$ and otherwise $1$. The test bits are then extracted with a bitwise $\&$ with $I$, producing the string $Z = z_10^mz_20^m\ldots z_m0^m$. This is almost what we want since $z_i = 1$ iff state $i$ is in $\Close_A(S)$. 
The final computation \emph{compresses} the $Z$ into the desired format. The multiplication produces the following length $2m^2$ string:
\begin{equation*}
\begin{split}
Z \times C &= z_10^mz_20^m\ldots z_m0^m \times 1(0^{m-1}1)^{m-1}  \\
&= z_10^{m-1} z_1z_20^{m-2} \cdots z_1\ldots z_{k}0^{m-k} \cdots z_1\ldots z_{m-1}0 z_1\ldots z_m  
0z_2\ldots z_m \cdots 0^{k}z_{k+1}\ldots z_m \cdots 0^{m-1} z_m 0^m  
\end{split} 
\end{equation*}
In particular, positions $m(m-1)+1$ through $m^2$ (from the left) contain the test bits compressed into a string of length $m$. The two shifts zeroes all other bits and moves this substring to the rightmost position in the word, producing the final result. Since $m = O(\sqrt{w})$ all of the above operations can be done in constant time.

Finally, observe that $\Insert_A$ and $\Member_A$ are trivially implemented in constant time. Thus, 
\begin{lemma}\label{lem:datastruct1}
For any TNFA with $m = O(\sqrt{w})$ states there is a simulation data structure using $O(m)$ space and $O(m^2)$ preprocessing time which supports all operations in $O(1)$ time. 
\end{lemma}   
The main bottleneck in the above data structure is the string $E$ that represents all $\epsilon$-paths. On a TNFA with $m$ states $E$ requires at least $m^2$ bits and hence this approach only works for $m = O(\sqrt{w})$. In the next section we show how to use the structure of TNFAs to do better.

\section{Overcoming the $\epsilon$-closure Bottleneck}\label{sec:notsimple}
In this section we show how to compute an $\epsilon$-closure on a TNFA with $m = O(w)$ states in $O(\log m)$ time. Compared with the result of the previous section we quadratically increase the size of the TNFA  at the expense of using logarithmic time. The algorithm is easily extended to an efficient simulation data structure. The key idea is a new hierarchical decomposition of TNFAs described below.

\subsection{Partial-TNFAs and Separator Trees} 
\begin{figure}[t]
  \centering \includegraphics[scale=.5]{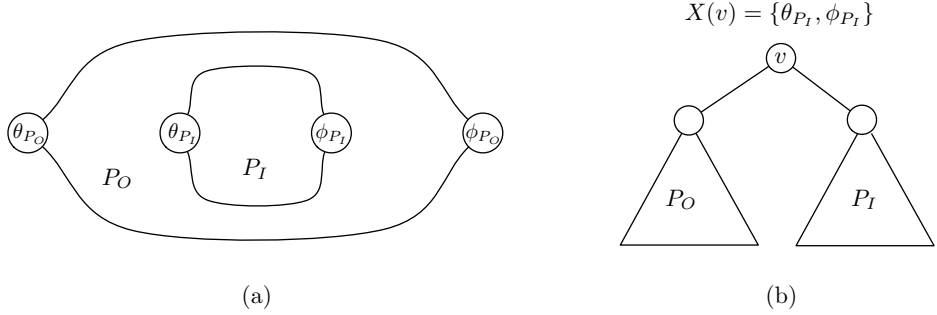}
   \caption{(a) Inner and outer pTNFAs. (b) The corresponding separator tree construction.}
   \label{fig:separator}
\end{figure}

First we need some definitions. Let $A$ be a TNFA with parse tree $T$. Each node $v$ in $T$ uniquely corresponds to two states in $A$, namely, the start and accepting states $\theta_{A'}$ and $\phi_{A'}$ of the TNFA $A'$ with the parse tree consisting of $v$ and all descendants of $v$. We say $v$ \emph{associates} the states $S(v) = \{\theta_{A'}, \phi_{A'}\}$. In general, if $C$ is a cluster of $T$, i.e., any connected subgraph of $T$, we say $C$ associates the \emph{set} of states $S(C) = \cup_{v \in C} S(v)$. We define the \emph{partial-TNFA} (pTNFA) for $C$, as the directed, labeled subgraph of $A$ induced by the set of states $S(C)$. In particular, $A$ is a pTNFA since it is induced by $S(T)$. The two states associated by the root node of $C$ are defined to be the start and accepting state of the corresponding pTNFA. We need the following result. 
\begin{lemma}\label{lem:separator}
For any pTNFA $P$ with $m>2$ states there exists a partitioning of $P$ into two subgraphs $P_O$ and $P_I$ such that
\begin{itemize}
  \item[(i)] $P_O$ and $P_I$ are pTNFAs with at most $2/3m + 2$ states each,
  \item[(ii)] any transition from $P_O$ to $P_I$ ends in $\theta_{P_I}$ and any transition from $P_I$ to $P_O$ starts in $\phi_{P_I}$, and
  \item[(iii)] the partitioning can be computed in $O(m)$ time. 
\end{itemize}
\end{lemma}
\begin{proof}
Let $P$ be pTNFA with $m > 2$ states and let $C$ be the corresponding cluster with $t$ nodes. Since $C$ is a binary tree with more than $1$ node, Jordan's classical result~\cite{Jordan1869} establishes that we can find in $O(t)$ time an edge $e$ in $C$ whose removal splits $C$ into two clusters each with at most $2/3t + 1$ nodes. These two clusters correspond to two pTNFAs, $P_O$ and $P_I$, and since $m = 2t$ each of these have at most $2/3m + 2$ states. Hence, (i) and (iii) follows. For (ii) assume w.l.o.g. that $P_O$ is the pTNFA containing the start and accepting state of $P$, i.e., $\theta_{P_O} = \theta_P$ and $\phi_{P_O} = \phi_P$. Then, $P_O$ is the pTNFA obtained from $P$ by removing all states of $P_I$. From Thompson's construction it is easy to check that any transition from $P_O$ to $P_I$ ends in $\theta_{P_I}$ and any transition from $P_I$ to $P_O$ must start in $\phi_{P_I}$.  \qed
\end{proof}
Intuitively, if we draw $P$, $P_I$ is "surrounded" by $P_O$, and therefore we will often refer to $P_I$ and $P_O$ as the \emph{inner pTNFA}  and the \emph{outer pTNFA}, respectively (see Fig.~\ref{fig:separator}(a)). 
Applying Lemma~\ref{lem:separator} recursively gives the following essential data structure. Let $P$ be a pTNFA with $m$ states. The \emph{separator tree} for $P$ is a binary, rooted  tree $B$ defined as follows: If $m=2$, i.e., $P$ is a trivial pTNFA consisting of two states $\theta_P$ and $\phi_P$, then $B$ is a single leaf node $v$ that stores the set $X(v) = \{\theta_P, \phi_P\}$. Otherwise ($m > 2$), compute $P_O$ and $P_I$ according to Lemma~\ref{lem:separator}. The root $v$ of $B$ stores the set $X(v) = \{\theta_{P_I}, \phi_{P_I}\}$, and the children of $v$ are roots of separator trees for $P_O$ and $P_I$, respectively (see Fig.~\ref{fig:separator}(b)). 

With the above construction each node in the separator tree naturally correspond to a pTNFA, e.g., the root corresponds to $P$, the children to $P_I$ and $P_O$, and so on. We denote the pTNFA corresponding to node $v$ in $B$ by $P(v)$. A simple induction combined with Lemma~\ref{lem:separator}(i) shows that if $v$ is a node of depth $k$ then $P(v)$ contains at most $(\frac{2}{3})^km+6$ states. Hence, the depth of $B$ is at most $d = \log_{3/2} m + O(1)$. By Lemma~\ref{lem:separator}(iii) each level of $B$ can be computed in $O(m)$ time and thus $B$ can be computed in $O(m\log m)$ total time.

\subsection{A Recursive $\epsilon$-Closure Algorithm}
We now present a simple $\epsilon$-closure algorithm for a pTNFA, which recursively traverses the separator tree $B$. We first give the high level idea and then show how it can be implemented in $O(1)$ time for each level of $B$. Since the depth of $B$ is $O(\log m)$ this leads to the desired result. For a pTNFA $P$ with $m$ states, a separator tree $B$ for $P$, and a node $v$ in $B$ define
\begin{relate}
\item[$\Close_{P(v)}(S)$:] 
\begin{enumerate}
\item Compute the set $Z \subseteq X(v)$ of states in $X(v)$ that are $\epsilon$-reachable from $S$ in $P(v)$.
\item If $v$ is a leaf return $S' := Z$, else let $u$ and $w$ be the children of $v$, respectively:
\begin{enumerate}
\item Compute the set $G \subseteq V(P(v))$ of states in $P(v)$ that are $\epsilon$-reachable from $Z$.
\item Return $S' := \Close_{P(u)}((S \cup G) \cap V(P(u))) \cup \Close_{P(w)}((S \cup G)\cap V(P(w)))$.
\end{enumerate}
\end{enumerate}
\end{relate}
\begin{lemma}\label{lem:correctness}
For any node $v$ in the separator tree of a pTNFA $P$, $\Close_{P(v)}(S)$ computes the set of states in $P(v)$ reachable via a path of $\epsilon$-transitions. 
\end{lemma}
\begin{proof}
Let $\widehat{S}$ be the set of states in $P(v)$ reachable via a path of $\epsilon$-transitions. We need to show that $\widehat{S} = S'$. It is easy to check that any state in $S'$ is reachable via a path of $\epsilon$-transitions and hence $S' \subseteq \widehat{S}$. We show the other direction by induction on the separator tree. If $v$ is leaf then the set of states in $P(v)$ is exactly $X(v)$. Since $S' = Z$ the claim follows. Otherwise, let $u$ and $w$ be the children of $v$, and assume w.l.o.g. that $X(v) =\{\theta_{P(u)}, \phi_{P(u)}\}$. Consider a path $p$ of $\epsilon$-transitions from state $s$ to state $s'$. There are two cases to consider:
\begin{description}
  \item[Case 1:] $s' \in V(P(u))$. If $p$ consists entirely of states in $P(u)$ then by induction it follows that $s' \in \Close_{P(u)}(S \cap V(P(u)))$. Otherwise, $p$ contain a state from $P(w)$. However, by Lemma~\ref{lem:separator}(ii) $\theta_{P(u)}$ is on $p$ and hence $\theta_{P(u)} \in Z$. It follows that $s' \in G$ and therefore $s' \in \Close_{P(u)}(G \cap V(P(u)))$.   
  \item[Case 2:] $s'\in V(P(w))$. As above, with the exception that $\phi_{P(u)}$ is now the state in $Z$. 
\end{description}
In all cases $s' \in S'$ and the result follows. \qed 
\end{proof}

\subsection{Implementing the Algorithm}
Next we show how to efficiently implement the above algorithm in parallel. The key ingredient is a compact mapping of states into positions in bitstrings. Suppose $B$ is the separator tree of depth $d$ for a pTNFA $P$ with $m$ states. The \emph{separator mapping} $M$ maps the states of $P$ into an interval of integers $[1, l]$, where $l = 3 \cdot 2^{d}$. The mapping is defined recursively according to the separator tree. Let $v$ be the root of $B$. If $v$ is a leaf node the interval is  $[1, 3]$. The two states of $P$, $\theta_{P}$ and $\phi_{P}$, are mapped to positions $2$ and $3$, respectively, while position $1$ is left intentionally unmapped. Otherwise, let $u$ and $w$ be the children of $v$. Recursively, map $P(u)$ to the interval $[1, l/2]$ and $P(w)$ to the interval $[l/2+1, l]$. Since the separator tree contains at most $2^d$ leaves and each contribute $3$ positions the mapping is well-defined. The size of the interval for $P$ is $l = 3 \cdot 2^{\log_{3/2} m + O(1)} = O(m)$. We will use the unmapped positions as test bits in our algorithm.

The separator mapping compactly maps all pTNFAs represented in $B$ into small intervals. Specifically, if $v$ is a node at depth $k$ in $B$, then $P(v)$ is mapped to an interval of size $l/2^k$ of the form $[(i-1)\cdot \frac{l}{2^k} + 1, i \cdot \frac{l}{2^k}]$, for some $1\leq i \leq 2^k$. The intervals that correspond to a pTNFA  $P(v)$ are \emph{mapped} and all other intervals are \emph{unmapped}. We will refer to a state $s$ of $P$ by its mapped position $M(s)$. A state-set of $P$ is represented by a bitstring $S$ such that, for all mapped positions $i$, $S[i] = 1$ iff the $i$ is in the state-set. Since $m = O(w)$, state-sets are represented in a constant number of words. 
 
To implement the algorithm we define a simple data structure consisting of four length $l$ bitstrings $X^\theta_k$, $X^\phi_k$, $E^\theta_k$, and $E^\phi_k$ for each level $k$ of the separator tree. For notational convenience, we will consider the strings at level $k$ as two-dimensional arrays consisting of $2^k$ intervals of length $l/2^k$, i.e., $X^\theta_k[i,j]$ is position $j$ in the $i$th interval of $X^\theta_k$. If the $i$th interval at level $k$ is unmapped then all positions in this interval are $0$ in all four strings. Otherwise, suppose that the interval corresponds to a pTNFA $P(v)$ and let $X(v) = \{\theta_v, \phi_v\}$. The strings are defined as follows: 
\begin{align*}
    X^\theta_k[i,j] =1 & \text{ iff $\theta_v$ is $\epsilon$-reachable in $P(v)$ from state $j$},  \\
    E^\theta_k[i,j] = 1 & \text{ iff state $j$ is $\epsilon$-reachable  in $P(v)$ from $\theta_v$}, \\
    X^\phi_k[i,j] = 1 & \text{ iff $\phi_v$ is $\epsilon$-reachable in $P(v)$ from state $j$},  \\  
   E^\phi_k[i,j] = 1 & \text{ iff state $j$ is $\epsilon$-reachable  in $P(v)$ from $\phi_v$}. 
\end{align*}
In addtion to these, we also store a string $I_k$ containing a test bit for each interval, that is, $I_k[i,j] = 1$ iff $j = 1$. Since the depth of $B$ is $O(\log m)$ the strings use $O(\log m)$ words. With a simple depth-first search they can all be computed in $O(m\log m)$ time.

Let $S$ be a bitstring representing a state-set of $A$. We implement the operation $\Close_A(S)$ by computing a sequence of intermediate strings $S_0, \ldots, S_d$ each corresponding to a level in the above recursive algorithm. Initially, $S_0 := S$ and the final string $S_d$ is the result of $\Close_A(S)$. At level $k$, $0 \leq k < d$,  we compute $S_{k+1}$ from $S_k$ as follows. Let $t = l/2^k - 1$.

\begin{align*}
Y^\theta &:= S_{k}\: \&\: X^\theta_k \\
Z^\theta &:= ((Y^\theta \: |\: I_k) - (I_k >> t)) \: \&\: I_k\\
F^\theta &:= Z^\theta - (Z^\theta >> t) \\
G^\theta &:= F^\theta \:\&\: E^\theta_k \\
Y^\phi &:= S_{k}\: \&\: X^\phi_k \\
Z^\phi &:= ((Y^\phi \: |\: I_k) - (I_k >> t)) \: \&\: I_k\\
F^\phi &:= Z^\phi - (Z^\phi >> t)\\
G^\phi &:= F^\phi \:\&\: E^\phi_k \\
S_{k+1} &:= S_k \:|\: G^\theta \:|\: G^\phi 
\end{align*}
We argue that the computation correctly simulates (in parallel) a level of the recursive algorithm. Assume that at the beginning of level $k$ the string $S_k$ represents the state-set corresponding the recursive algorithm after $k$ levels. We interpret $S_k$ as divided into $r = l/2^k$ intervals of length $t = l/2^k - 1$, each prefixed with a test bit, i.e., 
\begin{equation*}
S_k = 0s_{1,1}s_{1,2}\ldots s_{1,t}0s_{2,1}s_{2,2}\ldots s_{2,t}0 \ldots 0 s_{r,1}s_{r,2}\ldots s_{r,t} 
\end{equation*}
Assume first that all these intervals are mapped intervals corresponding to pTNFAs $P(v_1), \ldots, P(v_r)$, and let $X(v_i) = \{\theta_{v_i}, \phi_{v_i}\}$, $1\leq i \leq r$. Initially, $S_k\: \&\: X^\theta_k$ produces the string 
$$
Y^\theta = 0y_{1,1}y_{1,2}\ldots y_{1,t}0y_{2,1}y_{2,2}\ldots y_{2,t}0 \ldots 0 y_{r,1}y_{r,2}\ldots y_{r,t}, 
$$
where $y_{i,j} = 1$ iff $\theta_{v_i}$ is $\epsilon$-reachable in $P(v_i)$ from state $j$ and $j$ is in $S_k$. Then, similar to the second line in the simple algorithm, $(Y^\theta \: |\: I_k) - (I_k >> t) \: \&\: I_k$ produces a string of test bits $Z^\theta = z_10^tz_20^t \ldots z_r0^t$, where $z_i = 1$ iff at least one of $y_{i,1}\ldots y_{i,t}$ is $1$. In other words, $z_i = 1$ iff $\theta_{v_i}$ is $\epsilon$-reachable in $P(v_i)$ from any state in $S_k \cap V(P(v_i))$. Intuitively, the $Z^\theta$ corresponds to the "$\theta$-part" of the of $Z$-set in the recursive algorithm. Next we "copy" the test bits to get the string $F^\theta = Z^\theta - (Z^\theta >> t) = 0z_1^t0z_2^t\ldots 0z_r^t$. The bitwise $\&$ with $E^\theta_k$ gives
$$
G^\theta = 0g_{1,1}g_{1,2}\ldots g_{1,t}0g_{2,1}g_{2,2}\ldots g_{2,t}0 \ldots 0 g_{r,1}g_{r,2}\ldots g_{r,t}.
$$
By definition, $g_{i,j} = 1$ iff state $j$ is $\epsilon$-reachable in $P(v_i)$ from $\theta_{v_i}$ and $z_i = 1$. In other words, $G^\theta$ represents, for $1\leq i \leq r$, the states in $P(v_i)$ that are $\epsilon$-reachable from $S_k \cap V(P(v_i))$ through $\theta_{v_i}$. Again, notice the correspondance with the $G$-set in the recursive algorithm. The next $4$ lines are identical to first $4$ with the exception that $\theta$ is exchanged by $\phi$. Hence, $G^\phi$ represents the states that $\epsilon$-reachable through $\phi_{v_1}, \ldots, \phi_{v_r}$. 

Finally, $S_k \:|\: G^\theta \:|\: G^\phi$ computes the union of the states in $S_k$, $G^\theta$, and $G^\phi$ producing the desired state-set $S_{k+1}$ for the next level of the recursion. In the above, we assumed that all intervals were mapped. If this is not the case it is easy to check that the algorithm is still correct since the string in our data structure contain $0$s in all unmapped intervals. The algorithm uses constant time for each of the $d = O(\log m)$ levels and hence the total time is $O(\log m)$. 

\subsection{The Simulation Data Structure}
Next we show how to get a full simulation data structure. First, note that in the separator mapping the endpoints of the $\alpha$-transitions are consecutive (as in Sec.~\ref{sec:simple}). It follows that we can use the same algorithm as in the previous section to compute $\Move_A$ in $O(1)$ time. This requires a dictionary of bitstrings, $D_\alpha$, using additional $O(m)$ space and $O(m\log m)$ preprocessing time. The $\Insert_A$, and $\Member_A$ operations are trivially implemented in $O(1)$. Putting it all together we have:
\begin{lemma}\label{lem:datastruct2}
For a TNFA with $m = O(w)$ states there is a simulation data structure using $O(m)$ space and $O(m\log m)$ preprocessing time which supports all operations in $O(\log m)$ time.
\end{lemma}
Combining the simulation data structures from Lemmas~\ref{lem:datastruct1} and \ref{lem:datastruct2} with the reduction from Lemma~\ref{lem:simulation} and taking the best result gives Theorem~\ref{thm:main}. Note that the simple simulation data structure is the fastest when $m = O(\sqrt{w})$ and $n$ is sufficiently large compared to $m$.

\section{Remarks and Open Problems}\label{sec:remarks}
The presented algorithms assume a unit-cost multiplication operation. Since this operation is not in $AC^0$ (the class of circuits of polynomial size (in $w$), constant depth, and unbounded fan-in) it is interesting to reconsider what happens with our results if we remove multiplication from our machine model. The simulation data structure from Sec.~\ref{sec:simple} uses multiplication to compute $\Close_A$ and also for the constant time hashing to access $D_\alpha$. On the other hand, the algorithm of Sec.~\ref{sec:notsimple} only uses multiplication for the hashing. However, Lemma~\ref{lem:datastruct2} still holds since we can simply replace the hashing by binary search tree, which uses $O(\log m)$ time. It follows that Theorem~\ref{thm:main} still holds except for the $O(n + m^2)$ bound in the last line. 

Another interestring point is to compare our results with the classical Shift-Or algorithm by Baeza-Yates and Gonnet~\cite{BYG1992} for exact pattern matching. Like ours, their algorithm simulates a NFA with $m$ states using word-level parallelism. The structure of this NFA permits a very efficient simulation with an $O(w)$ speedup of the simple $O(nm)$ time simulation. Our results generalize this to regular expressions with a slightly worse speedup of $O(w/\log w)$. We wonder if it is possible to remove the $O(\log w)$ factor separating these bounds. 

From a practical viewpoint, the simple algorithm of Sec.~\ref{sec:simple} seems very promising since only about $15$ instructions are needed to carry out a step in the state-set simulation. Combined with ideas from~\cite{NR2004} we believe that this could lead to a practical improvement over previous algorithms.

\section{Acknowledgments}
The author wishes to thank Rasmus Pagh and Inge Li G{\o}rtz for many comments and interesting discussions.
\bibliographystyle{abbrv}
\bibliography{paper}

\end{document}